\documentclass{article}
\usepackage{caption}
\usepackage{wrapfig}
\usepackage{float}
\usepackage[margin=1.5in, includeheadfoot]{geometry}
\usepackage[backend=bibtex,style=numeric]{biblatex}
\usepackage[utf8]{inputenc} 
\usepackage[T1]{fontenc}    
\usepackage{hyperref}       
\usepackage{url}            
\usepackage{booktabs}       
\usepackage{amsfonts}       
\usepackage{nicefrac}       
\usepackage{microtype}      
\usepackage{lipsum}
\usepackage{fancyhdr}       
\usepackage{graphicx}       
\graphicspath{{media/}}     

\title{Beyond single-channel agentic benchmarking }

\author{Nelu D. Radpour \\
  Florida State University \\
 radpour@psy.fsu.edu} 
 \date{} %
\begin{document}
\maketitle

\begin{abstract}
Contemporary benchmarks for agentic artificial intelligence (AI) frequently evaluate safety through isolated task-level accuracy thresholds \cite{ren2024safetywashing, li2024scisafeeval}, implicitly treating autonomous systems as single points of failure. This single-channel paradigm diverges from established principles in safety-critical engineering, where risk mitigation is achieved through redundancy, diversity of error modes, and joint system reliability. This paper argues that evaluating AI agents in isolation systematically mischaracterizes their operational safety when deployed within human-in-the-loop environments. Using a recent laboratory safety benchmark \cite{zhou2026benchmarking} as a case study demonstrates that even imperfect AI systems can nonetheless provide substantial safety utility by functioning as redundant audit layers against well-documented sources of human failure, including vigilance decrement, inattentional blindness, and normalization of deviance. This perspective reframes agentic safety evaluation around the reliability of the human-AI dyad rather than absolute agent accuracy, with a particular emphasis on uncorrelated error modes as the primary determinant of risk reduction. Such a shift aligns AI benchmarking with established practices in other safety-critical domains and offers a path toward more ecologically valid safety assessments.
\end{abstract}


\section{Introduction}

The evaluation of agentic AI systems increasingly relies on benchmark-driven assessments that reduce safety to task-level performance metrics, often expressed as fixed accuracy thresholds. While such benchmarks offer standardization and comparability, they implicitly assume a single-channel operational model in which the AI system functions as an authoritative decision-maker. This assumption stands in tension with decades of research in safety engineering, cognitive ergonomics, and human factors, where risk mitigation is achieved not through perfect agents but through layered, redundant systems designed to tolerate individual failure. While prior work has examined AI as automated judges, multi-agent coordination, and human-AI teaming, these literatures typically evaluate agent competence or interaction quality in isolation. In contrast, this paper frames safety as an emergent property of joint human-AI reliability, with uncorrelated error modes rather than individual accuracy as the primary determinant of risk reduction.

In a recent article, Zhou et al. \cite{zhou2026benchmarking} introduced LabSafety Bench to evaluate the reliability of both large language models (LLMs) and vision language models (VLMs) in scientific laboratories. They conclude that current models are unsafe for deployment, particularly for tasks like experimental design and autonomous orchestration, because they fail to surpass a 70\% accuracy threshold in hazard identification. While the creation of this benchmark is a notable contribution, the conclusion drawn rests on a misconception of risk mitigation frameworks as they pertain to laboratory safety. The same authors go on to cite the 2023 GMFC Labs explosion as evidence that flawed model recommendations could escalate accidents. However, by evaluating AI agents in isolation rather than as redundant layers in a human system, the study fails to account for the primary driver of laboratory accidents, which is human variability. This paper posits that a consistent, albeit imperfect, AI agent offers a critical safety margin against the stochastic lapses such as fatigue and inattentional blindness \cite{simons1999gorillas} that define human performance.

\section{Related Work and Benchmarking Paradigms}

A growing body of research has examined human-AI collaboration, AI-assisted decision making, and multi-agent systems across domains such as medicine, emergency response, and complex reasoning. Across these settings, joint human-AI performance has been shown to depend not only on agent accuracy, but on complementary strengths, trust calibration, and the interaction of distinct failure modes \cite{parasuraman2010complacency, vaccaro2024combinations, liu2025human}. These findings consistently indicate that combined systems may outperform either humans or AI alone, even when individual components remain imperfect, provided that their errors are not strongly correlated.

Within this broader landscape, adjacent work has explored the use of AI systems as automated evaluators or judges, particularly in the benchmarking and assessment of LLMs. These approaches aim to scale evaluation, reduce human variance, or provide consistent scoring of model outputs. Studies in this area have demonstrated that complementary capabilities can yield joint outcomes that exceed the performance of either actor alone by leveraging distinct error profiles and appropriately calibrated trust \cite{schemmer2023towards}. For example, work in complex diagnostic tasks shows that human experts and algorithmic assistants may achieve superior joint accuracy when each contributes distinct strengths, with performance shaped by user expertise and trust dynamics \cite{zhang2020effect}. Similarly, in domains such as medical bias assessment, structured forms of human-AI collaboration have been shown to enhance reliability and reduce intervention rates relative to standalone evaluations \cite{reverberi2022experimental}. Collectively, this literature underscores the importance of modeling interaction dynamics rather than evaluating agent competence in isolation.

Despite these advances, much of the AI benchmarking literature continues to emphasize isolated accuracy metrics, task-level performance, or interaction quality without systematically framing safety as a function of joint reliability under uncorrelated error modes. In many benchmarks, AI systems are implicitly modeled as replacements for human judgment within the evaluation pipeline rather than as safety-relevant agents embedded in operational workflows. As a result, benchmark outcomes are often interpreted as absolute indicators of system-level safety or risk, despite the absence of any explicit modeling of how human and AI errors interact in practice. This pattern is particularly salient in recent safety benchmarks for agentic and laboratory-facing AI systems, which frequently report sub-threshold model performance as evidence of categorical unsuitability for deployment.

This paper reframes agentic safety benchmarking around joint reliability in human-AI systems. Rather than evaluating whether an AI agent meets a predefined accuracy threshold in isolation, we emphasize redundancy and error diversity as the primary determinants of risk reduction in high-stakes environments. Existing benchmarking approaches do not explicitly operationalize this perspective for safety evaluation in agentic AI systems.

\section{Safety as Joint Human-AI Reliability}
In the realm of safety engineering, tools are not evaluated in isolation, but by their ability to close potential holes in existing defensive layers. Decades of research in cognitive ergonomics and occupational safety demonstrate that human performance in high-stakes environments is anything but flawless \cite{ reason1990contribution}. It is inherently stochastic and degrades unpredictably under the influence of fatigue, stress, and circadian rhythm disruption. This phenomenon, known as the vigilance decrement, means that a human researcher who is safe at 9:00 a.m. may be significantly less safe at 9:00 p.m \cite{warm2008vigilance}. 

From a systems reliability perspective, safety is determined by the joint probability of failure across interacting agents rather than the standalone accuracy of any single component. In a human-only workflow, hazards go undetected when cognitive lapses coincide with task demands. In a human-AI workflow, system-level failure occurs only when both the human and the AI miss the same hazard. When human and AI error modes are weakly correlated, even agents with modest standalone accuracy can substantially reduce overall failure probability. This distinction between marginal performance and joint reliability is well established in safety engineering but remains largely absent from agentic AI benchmarking.

When Zhou et al. \cite{zhou2026benchmarking} warn that flawed recommendations by AI during experimental design could lead to disasters like the 2023 GMFC Labs explosion (an incident where static energy build-up was overlooked), they inadvertently highlight the necessity of algorithmic redundancy. While the precise cause of that tragedy was undoubtedly complex, it serves as a stark reminder that human safety systems are not impervious to conventional risk factors. 
In safety engineering, the goal is not to find a single perfect monitor, but to create overlapping layers of defense in what is often described as the Swiss cheese model \cite{reason1990contribution, reason2006revisiting}. A shift toward measuring joint reliability is already yielding significant gains in other safety-critical fields. For example, recent prospective cohort studies in clinical medicine have demonstrated that the integration of AI decision support into human workflows reduced diagnostic error rates from 22 to 12\%, representing a 45\% reduction in systemic failure \cite{liu2025human}. This finding confirms that a human plus AI parallel system achieves a net reliability higher than a human in isolation, particularly when the AI acts as a simultaneous audit layer rather than a sequential replacement. This phenomenon is further supported by findings in emergency medicine, where AI-driven tools have been shown to reduce misdiagnoses by 37\% in high pressure settings \cite{taylor2025leveraging}. By providing a redundant check that captures the stochastic lapses of human attention, such tools create a safety net that functions effectively even when individual agents are imperfect.

\begin{wrapfigure}{r}{0.6\textwidth}
  \centering
  \includegraphics[width=\linewidth]{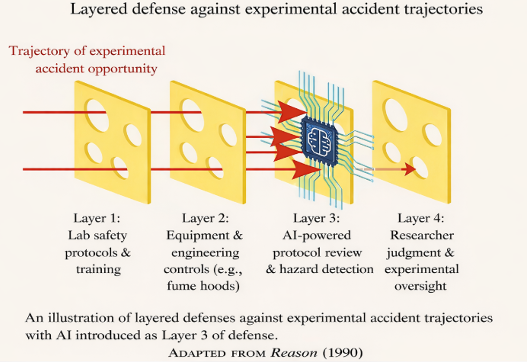}
   {\textbf{Figure 1}: AI-Enhanced Swiss Cheese Model}
  
\end{wrapfigure}

Figure 1 modifies Reason’s original Swiss cheese model \cite{reason1990contribution, reason2006revisiting} to illustrate how an AI layer adds value not through exceptional insight, but because its error modes are entirely distinct from human error modes. A language model does not suffer from biological fatigue or experience tunnel vision during an emergency, nor does it normalize deviation after months of routine work. 

If an unassisted human researcher identifies 80\% of hazards and an AI assistant identifies a different and overlapping 70\%, the joint probability of a safety failure decreases significantly. While the realization of this gain depends on AI adherence (the human’s willingness to cross-reference and act upon the agent's prompt, a factor influenced by the documented risks of automation complacency \cite{parasuraman2010complacency}, arbitrary dismissal of 60 to 70\% accuracy as "unsafe" ignores the potential for cumulative system reliability. As demonstrated in recent meta-analyses of human-AI synergy \cite{vaccaro2024combinations}, the objective of teaming is not for the agent to achieve superhuman autonomy, but to provide a complementary layer that outperforms the human alone. A tool that catches two out of three hazards is operationally useless only if we assume the human operator catches nothing; however, if we accept that the human is competent but fallible, AI serves as a high-volume and low-latency audit mechanism that catches what slips through the cracks of human attention. A few instances of how these joint probabilities would impact the net safety gain are presented in Table 1 below.

\begin{table}[H]
  \centering
    \includegraphics[width=0.8\textwidth]{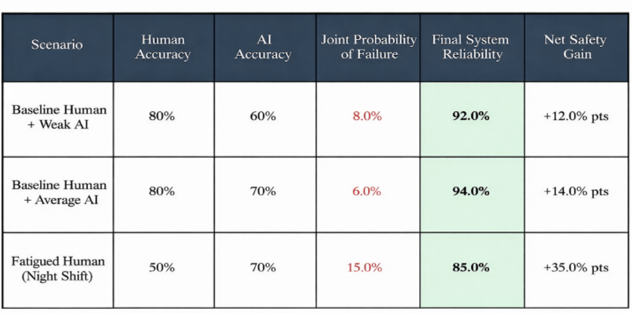}
    \caption{Projected impact of AI redundancy on net system reliability}
\end{table}

Moreover, the survey data provided by Zhou et al. offer a critical empirical foundation for this perspective. The report that over 30\% of surveyed researchers expressed moderate to high trust in current model suggestions is framed as evidence of a dangerous illusion of understanding. However, when situated within the context of contemporaneous cognitive modeling, such as the Centaur model, a framework designed to predict human internal states and decision-making patterns \cite{binz2025foundation}, this statistic reveals the depth of the human element as the primary variable of risk. If we accept the human researcher as a biologically fallible agent, the observed variability in trust represents a transition toward deferred reliance in high-pressure environments. 

This shift is often a necessary response to the illusions of competence where experts systematically overestimate their own real-time error detection capabilities \cite{koriat2005illusions}. Trust in AI systems frequently emerges as a strategic hedge against this recognized human fallibility \cite{galindez2025trust}.
This inherent propensity for inattentional blindness remains the most pervasive risk in physical laboratories, as intense focus on primary tasks frequently occludes peripheral hazards \cite{mack2003inattentional}. As demonstrated in expert observers who consistently fail to detect salient anomalies while performing routine diagnostic tasks, even highly trained individuals are not immune to these cognitive lapses \cite{drew2013invisible}. Consequently, even a clumsy or verbose suggestion from an artificial agent can disrupt a dangerous error chain by forcing a “cognitive pause,” prompting a reevaluation of the environment. While we must distinguish this interruptive utility from the broader risks of emergent misalignment \cite{betley2026training}, where narrow training produces deceptive or broadly harmful behaviors in unrelated domains, the silence of a missed hazard remains the far more consequential failure in high-risk settings. For hazards like static energy build-up, a verbose LLM acting as a heuristic prompt is a critical safeguard. In this hierarchy of risk, a false positive is an operational friction, whereas a false negative is a systemic catastrophe.

Failure modes in human-AI systems can be broadly categorized into human omission errors, AI commission errors, and correlated failures arising from over-reliance or automation complacency. Human omission errors dominate safety-critical environments due to attentional limits, fatigue, and normalization of deviance \cite{nushi2018towards, mueller2021principles}. AI commission errors more often manifest as false positives or verbose warnings that introduce operational friction rather than catastrophic risk. The most dangerous regime arises when errors become correlated, such as when humans defer uncritically to incorrect AI outputs. When AI systems function as advisory or audit layers, however, their errors remain largely uncorrelated with human attentional failures, producing a net gain in system safety despite imperfect agent accuracy.

Evaluating safety in agentic AI systems therefore requires benchmarks that move beyond isolated accuracy thresholds toward metrics that capture joint detection probability, error correlation, and human adherence. A system that identifies hazards at subthreshold rates may nonetheless provide substantial safety benefit if its detections are orthogonal to human blind spots, particularly in domains where false negatives carry disproportionately high costs. Without explicitly modeling these interactions, benchmark results risk being misinterpreted as absolute indicators of system safety rather than as components of a broader reliability analysis.

While LabSafety Bench provides a valuable standardized evaluation, this paper offers a perspective based on parallel system reliability to demonstrate that future assessments must move beyond isolated accuracy thresholds to measure the net safety utility of the human-AI dyad. Reducing complex safety dynamics to a decontextualized 70\% accuracy threshold risks mischaracterizing the operational safety of a system and misinforming readers who may interpret such figures as an absolute measure of danger rather than a single variable in a larger equation. We hope that future work will enable the variables of model reliability and human adherence to be decoupled and more deeply understood.

\section{Conclusion}
In safety-critical settings, the most consequential failures are not false positives but silent omissions. Human operators, even when highly trained, are vulnerable to attentional lapses, fatigue, and normalization of deviance. Within this context, an imperfect but persistent AI audit layer can provide a meaningful safety margin by interrupting error chains and prompting reevaluation at moments when human cognition is most brittle.

Benchmarks that classify systems as safe or unsafe based solely on isolated task performance risk misinforming both researchers and practitioners (and the general public) by conflating agent accuracy with system reliability. A more ecologically valid approach would quantify the safety margin delta introduced by AI assistance, explicitly modeling human adherence, trust calibration, and joint error probabilities.

The central question, therefore, is not whether an AI agent is safe in isolation, but how much safer a sociotechnical system becomes when that agent is introduced as a redundant layer. Aligning AI benchmarking with this principle is essential if safety evaluations are to reflect real-world operational risk rather than abstract performance ceilings.

\printbibliography

\end{document}